\documentclass[aps,prb,twocolumn,floatfix]{revtex4}
\usepackage{graphicx}
\usepackage{latexsym}
\usepackage{amsmath}
\usepackage{graphics}
\usepackage{amssymb}
\usepackage{layout}
\usepackage{verbatim}
\usepackage{amsfonts,epsfig}
\newcommand{\ket}[1]{\left|#1\right>}
\newcommand{\bra}[1]{\left< #1 \right|}

\begin{document}

\title{Energy relaxation of a superconducting charge qubit via Andreev processes}

\author{R. M. Lutchyn and L. I. Glazman}

\affiliation{ W.I.\ Fine Theoretical Physics Institute, University
of Minnesota, Minneapolis, Minnesota 55455, USA}

\date{\today }

\begin{abstract}
  We study fundamental limitations on the energy relaxation rate of a
  superconducting charge qubit with a large-gap Cooper-pair box, $\Delta_b > \Delta_r$.
  At a sufficiently large mismatch between the gap energies in the box $\Delta_b$ and in the reservoir $\Delta_r$,
  ``quasiparticle poisoning" becomes ineffective even in the presence
  of nonequilibrium quasiparticles in the reservoir. The qubit
  relaxation still may occur due to higher-order (Andreev) processes.
In this paper we evaluate the qubit energy relaxation rate
$T_1^{-1}$ due to Andreev processes.
\end{abstract}
\maketitle

\section{Introduction}
A large number of recent experimental studies~\cite{Mannik,
Aumentado, Ferguson, Turek, Yamamoto, Guillaume, Zorin,
Gunnarsson, Naaman} indicates the presence of quasiparticles in
superconducting single-charge devices at low temperatures. The
operation of these devices, of which the best known is Cooper-pair
box qubit, requires $2e$-periodic dependence of the charge of the
box on its gate voltage, and thus, an introduction of an unpaired
electron(quasiparticle) in the Cooper-pair box (CPB) is a
significant problem. The superconducting charge qubit operates at
the degeneracy point for Cooper-pairs, $N_g=1$, with $N_g$ being
the dimensionless gate voltage. For equal gap energies in the
Cooper-pair box and reservoir, $\Delta_b=\Delta_r$, the states of
the qubit at $N_g=1$ are unstable with respect to quasiparticle
tunneling to the box. The quasiparticle changes the charge state
of CPB from even to odd, and lowers the charging energy. This
phenomenon, commonly referred to as ``quasiparticle poisoning'',
is well-known from the studies of the charge parity effect in
superconductors, see, for example, Matveev \emph{et.
al.}~\cite{Matveev} and references therein. ``Quasiparticle
poisoning" can degrade the performance of the charge qubit in two
ways. First, it causes the operating point of the qubit to shift
stochastically on the time scale comparable with the measurement
time~\cite{Naaman}. Second, it contributes to the
decoherence~\cite{Lutchyn2006}. One of the approaches to improve
the performance of charge qubits is to use superconducting gap
engineering. In most single-charge superconducting devices
``quasiparticle poisoning'' can be suppressed even in the presence
of nonequilibrium quasiparticles in the reservoir by engineering a
large mismatch between $\Delta_b$ and $\Delta_r$. Gap energies in
superconductors can be modified by oxygen doping~\cite{Aumentado},
applying a magnetic field~\cite{Turek, Gunnarsson}, and adjusting
layer thickness~\cite{Yamamoto, Ferguson}. In this paper we study
the fundamental limitations on the energy relaxation time in a
charge qubit with a large gap in the box, $\Delta_b>\Delta_r$.

\begin{figure}
\centering
\includegraphics[width=2.8in]{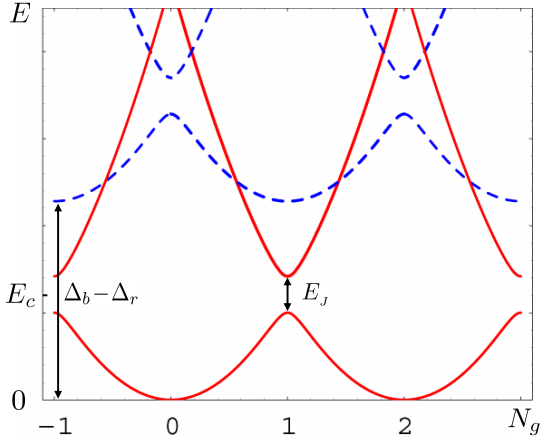}
\caption{(color online). The spectrum of the Cooper-pair box as a
function of the dimensionless gate voltage for a large-gap
mismatch, $\Delta_b > \Delta_r$. The solid and dashed lines
correspond to an even- and odd-charge state of the box,
respectively.}\label{fig5-1}
\end{figure}

For equal gap energies in the box and reservoir,
$\Delta_b=\Delta_r$, the energy relaxation rate due to
``quasiparticle poisoning"~\cite{Lutchyn2006} is
\begin{equation}\label{T1_old}
\frac{1}{T_1}\propto\frac{g_{_T}n_{\rm{qp}}}{\hbar\nu_F}\sqrt{\frac{T}{E_{_J}}}
\end{equation} with $n_{\rm{qp}}$, $g_{_T}$ and $\nu_F$ being the density of
quasiparticles in the reservoir, dimensionless conductance of the
junction and density of states at the Fermi level, respectively.
The relaxation rate $1/T_1$ in Eq.~(\ref{T1_old}) was derived
under the assumption that an unpaired electron tunnels from the
reservoir to the box to minimize the energy of the system. Indeed,
for $\Delta_b=\Delta_r$, the odd-charge state of the CPB has lower
energy at $N_g=1$ due to the Coulomb blockade effect. By properly
engineering superconducting gap energies (\emph{i.e.} inducing
large gap mismatch, $\Delta_b>\Delta_r$), one can substantially
reduce quasiparticle tunneling rate to the Cooper-pair box.
Suppose initially the qubit is in the excited state with energy
$E_{\ket{+}}$, and the quasiparticle is in the reservoir with
energy $E_p$. Upon quasiparticle tunneling to the box, the minimum
energy of the final state is $E^{\rm {min}}_f=\Delta_b+E_{N+1}$
with $E_{N+1}$ being the energy of the CPB in the odd-charge
state. Therefore, the threshold energy for a quasiparticle to
tunnel to the box is $E^{\rm min}_p=\Delta_b+E_{N+1}-E_{\ket{+}}$,
see also Fig.~\ref{fig5-1}. If $E^{\rm min}_p-\Delta_r\gtrsim
E_{_J}\gg T$, only exponentially small fraction of quasiparticles
are able to tunnel into the island. (Note that the energy
difference between excited and ground state of a charge qubit is
$E_{_J}$, while the energy of the qubit in the excited state is
$E_{\ket{+}}=E_c+E_{_J}/2$. Here $E_c$, $E_{_J}$ and $T$ are the
charging energy of the CPB, the Josephson energy associated with
the tunnel junction, and the temperature, respectively.) Thus, the
contribution to the qubit relaxation rate $T_1^{-1}$ from the
processes involving real quasiparticle tunneling to the island
becomes
\begin{eqnarray}\label{T1_exp}
\frac{1}{T_1} \propto
\frac{g_{_T}n_{\rm{qp}}}{\hbar\nu_F}\exp\left(-\frac{\Delta_b\!-\!\Delta_r
\!-\! E_c \!-\! E_{_J}/2}{T}\right),
\end{eqnarray}
and is much smaller than the one of Eq.~(\ref{T1_old}). (To obtain
Eq.~(\ref{T1_exp}), we used the fact that $E_{N+1}=0$ at $N_g=1$.)
However, there is also a mechanism of energy relaxation
originating from the higher order tunneling processes (Andreev
reflection). The contribution of these processes to the qubit
relaxation is activationless, and can be much larger than the one
of Eq.~(\ref{T1_exp}). In the rest of the paper we study qubit
energy relaxation due to Andreev processes in detail.

\section{Theoretical model}
Dynamics of the Cooper-pair box coupled to the superconducting
reservoir through the tunnel junction is described by the
Hamiltonian
\begin{eqnarray}\label{Hqubit}
H=H_{_C}\!+\!H^b_{\rm{BCS}}\!+\!H^r_{\rm{BCS}}\!+\!H_{_T}.
\end{eqnarray}
Here $H^b_{\rm{BCS}}$ and $H^r_{\rm{BCS}}$ are BCS Hamiltonians
for the box and reservoir; $H_{_C}=E_c(Q/e\!-\!N_g)^2$ with $N_g$
and $Q$ being the dimensionless gate voltage and the charge of the
CPB, respectively. We consider the following energy scale
hierarchy: $\Delta_b> \Delta_r, E_c> E_{_J} \gg T$. In order to
distinguish between Cooper-pair and quasiparticle tunneling, we
present the Hamiltonian~(\ref{Hqubit}) in the
form~\cite{Lutchyn2005}
\begin{eqnarray}\label{Hqubit2}
H=H_0\!+\!V,\mbox{ and }V=H_{_T}\!-\!H_{_{_J}}.
\end{eqnarray}
Here
$H_0=H_{_C}\!+\!H^b_{\rm{BCS}}\!+\!H^r_{\rm{BCS}}\!+\!H_{_J}$, and
$H_{_J}$ is the Hamiltonian describing Josephson tunneling
\begin{eqnarray}
H_{_J}=\ket{N}\bra{N} H_{_T} \frac{1}{E\!-\!H_0}H_{_T}
\ket{N\!+2}\bra{N\!+2} \!+\!{\rm{H.c.}}\nonumber
\end{eqnarray}
The matrix element $\bra{N} H_{_T} \frac{1}{E\!-\!H_0}H_{_T}
\ket{N\!+\!2}$ is proportional to the Josephson energy $E_{_J}$.
The perturbation Hamiltonian $V$ defined in Eq.~(\ref{Hqubit2}) is
suitable for calculation of the quasiparticle tunneling rate. The
tunneling Hamiltonian for homogeneous insulating barrier is
\begin{eqnarray}\label{tunneling}
H_T=\sum_{\sigma}\int{d\mathbf{x}d\mathbf{x'}}\left(T(\mathbf{x},\mathbf{x'})\Psi^{\dag}_{\sigma}(\mathbf{x})\Psi_{\sigma}(\mathbf{x'})\!+\!{\rm{H.c.}}\right),
\end{eqnarray}
where $\mathbf{x}$ and $\mathbf{x'}$ denote the coordinates in the
CPB and reservoir, respectively, and $T(\mathbf{x},\mathbf{x'})$,
in the limit of a barrier with low transparency, is defined as
\begin{eqnarray}\label{tunn}
T(\mathbf{x},\mathbf{x'})=\frac{1}{4\pi^2}\sqrt{\frac{\mathcal{T}}{\nu_F^2}}\delta^2(\mathbf{r}\!-\!\mathbf{r'})\delta(z)\delta(z')\frac{\partial}{\partial
z}\frac{\partial}{\partial z'}.
\end{eqnarray}
Here $\mathcal{T}$ is the transmission coefficient of the barrier,
$\mathbf{r}$ and $z$ are the coordinates in the plane of the
tunnel junction and perpendicular to it, respectively. The
Hamiltonian~(\ref{tunneling}) along with the above definition of
$T(\mathbf{x},\mathbf{x'})$ properly takes into account the fact
that in the tunnel-Hamiltonian approximation the wavefunctions
turn to zero at the surface of the junction~\cite{PradaSols,
Houzet}. In terms of the transmission coefficient $\mathcal{T}$,
the dimensionless conductance of the tunnel junction $g_{_T}$  can
be defined as
$g_{_T}=\mathcal{T}S_Jk_F^2/4\pi=\frac{1}{3}\mathcal{T}N_{\rm
ch}$, where $S_J$ is the area of the junction, and $N_{\rm ch}$ is
the number of transverse channels in the junction.

The energy relaxation rate of the qubit due to higher-order
processes is given by
\begin{eqnarray}\label{rate}
\Gamma_{A}\! =\!\frac{2\pi}{\hbar}\sum_{p,p'}\!&\!2\!&\!|A_{p'p}
|^2\delta(E_{p'}\!-\!E_p\!-\!E_{_J})f_{_F}(E_p)(1\!-\!f_{_F}(E_{p'})).\nonumber\\
\end{eqnarray}
Here $f_{_F}(E_p)$ is the Fermi distribution function with
$E_p=\sqrt{\varepsilon_p^2\!+\!\Delta_r^2}$ being the energy of a
quasiparticle in the reservoir. The amplitude $A_{p'p}$ is given
by the second order perturbation theory in $V$,
\begin{eqnarray}\label{AM0}
A_{p'p}
=\bra{-,E_{p'\uparrow}}V\frac{1}{E_i\!-\!H_0}V\ket{+,E_{p\uparrow}}.\\\nonumber
\end{eqnarray}
At $E_c\gg E_{_J}$ and $N_g=1$, the eigenstates of the qubit are
given by the symmetric and antisymmetric superposition of two
charge states, \emph{i.e.}
$\ket{-}=\frac{\ket{N}+\ket{N+2}}{\sqrt{2}}$ and
$\ket{+}=\frac{\ket{N}-\ket{N+2}}{\sqrt{2}}$ with the
corresponding eigenvalues $E_{\ket{\pm}}=E_c\pm E_{_J}/2$. In the
initial moment of time the qubit is prepared in the excited state
and the quasiparticle is in the reservoir, \emph{i.e}
$\ket{+,E_{p\uparrow}}\equiv\ket{+}\otimes\ket{E_{p\uparrow}}$.
The energy of the initial state is $E_i=E_p+E_{\ket{+}}$. The
denominator in the amplitude~(\ref{AM0}) corresponds to the
formation of the virtual intermediate state when the quasiparticle
has tunnelled to the island from the reservoir. Since a
quasiparticle is a superposition of a quasi-electron and
quasi-hole, the contributions to $A_{p'p}$ come from two
interfering paths:
\begin{eqnarray}\label{AM}
A_{p'p} &=&\frac{1}{2}
\bra{N\!+\!2,E_{p'\uparrow}}V\frac{1}{E_i\!-\!H_0}V\ket{N,E_{p\uparrow}}\nonumber\\\nonumber\\
&\!-\!&\frac{1}{2}\bra{N,E_{p'\uparrow}}V\frac{1}{E_i\!-\!H_0}V\ket{N\!+\!2,E_{p\uparrow}}.
\end{eqnarray}
To calculate the amplitude $A_{p' p}$, we use particle-conserving
Bogoliubov transformation~\cite{Schrieffer, Bardeen, Josephson}:
\begin{eqnarray}\label{particleconserving}
\gamma^{\dag}_{n
\sigma}&=&\int\!{d\mathbf{x}}\left[U_{n}(\mathbf{x})\Psi^{\dag}_{\sigma}(\mathbf{x})-\sigma
V_{n}(\mathbf{x})\Psi_{-\sigma}(\mathbf{x})R^{\dag}\right]\nonumber\\
\gamma_{n
\sigma}&=&\int\!{d\mathbf{x}}\left[U_{n}(\mathbf{x})\Psi_{\sigma}(\mathbf{x})-\sigma
V_{n}(\mathbf{x})\Psi^{\dag}_{-\sigma}(\mathbf{x})R\right]
\end{eqnarray}
The operators $R^{\dag}$ and $R$ transform a given state in an
$N$-particle system into the corresponding state in the $N+2$ and
$N-2$ particle system, respectively, leaving the quasiparticle
distribution unchanged, \emph{i.e.}
$R^{\dag}\ket{N}=\ket{N\!+\!2}$. Thus, quasiparticle operators
$\gamma^{\dag}_{n \sigma}$ and $\gamma_{n \sigma}$ defined in
Eq.~(\ref{particleconserving}) do conserve particle
number~\cite{metallic}. The transformation coefficients
$U_n(\mathbf{x})$ and $V_n(\mathbf{x})$ are given by the solution
of Bogoliubov-de Gennes equation. For spatially homogenous
superconducting gap $\Delta$, the functions $U_{n}(\mathbf{x})$
and $V_{n}(\mathbf{x})$ can be written as
$U_{n}(\mathbf{x})=u_n\phi_n(\mathbf{x})$ and
$V_{n}(\mathbf{x})=v_n\phi_n(\mathbf{x})$. The coherence factors
$u_n$ and $v_n$ are given by
\begin{eqnarray}
u_n^2=\frac{1}{2}\left(1\!+\!\frac{\varepsilon_{n}}{E_n}\right)
\mbox{ and }
v_n^2=\frac{1}{2}\left(1\!-\!\frac{\varepsilon_{n}}{E_n}\right).\nonumber
\end{eqnarray}
Here $E_n=\sqrt{\varepsilon_{n}^2+\Delta^2}$; $\varepsilon_{n}$
and $\phi_n(\mathbf{x})$ are exact eigenvalues and eigenfunctions
of the single-particle Hamiltonian, which may include random
potential $\mathcal{V}(\mathbf{x})$, \emph{e.g.}, due to
impurities. The single-particle energies $\varepsilon_{n}$ and
wavefunctions $\phi_n(\mathbf{x})$ are defined by the following
Shr\"{o}dinger equation:
\begin{eqnarray}
\left[-\frac{\hbar^2}{2m}\vec{\nabla}^2+\mathcal{V}(\mathbf{x})\right]\phi_n(\mathbf{x})=\varepsilon_{n}\phi_n(\mathbf{x}).\nonumber
\end{eqnarray}
In the presence of time-reversal symmetry $u_{n}$, $v_{n}$ and
$\phi_n(\mathbf{x})$ can be taken to be real. Then with the help
of Eq.~(\ref{particleconserving}), we obtain the amplitude of the
process $A_{p' p}$:
\begin{widetext}
\begin{eqnarray}\label{A1}
A_{p'
p}=\frac{1}{2}\int{d\mathbf{\mathbf{x_1}}d\mathbf{x_1'}d\mathbf{x_2}d\mathbf{x_2'}}T(\mathbf{x_1},\mathbf{x_1'})T(\mathbf{x_2},\mathbf{x_2'})\left[U_{p'}(\mathbf{x_1'})V_{p}(\mathbf{x_2'})\!-\!U_{p}(\mathbf{x_1'})V_{p'}(\mathbf{x_2'})\right]\sum_{k}\frac{U_{k}(\mathbf{x_1})V_{k}(\mathbf{x_2})}{E_p\!+\!\delta
E_{+}\!-\!E_k},
\end{eqnarray}
\end{widetext}
where $\delta E_{+}\equiv E_{\ket{+}}-E_{N+1}=E_c+E_{_J}/2$. The
minus sign in the parenthesis here reflects the destructive
interference between quasi-electron and quasi-hole contributions,
see also Eq.~(\ref{AM}).

\section{Disorder averaging}
\begin{figure}
\centering
\includegraphics[width=3.2in]{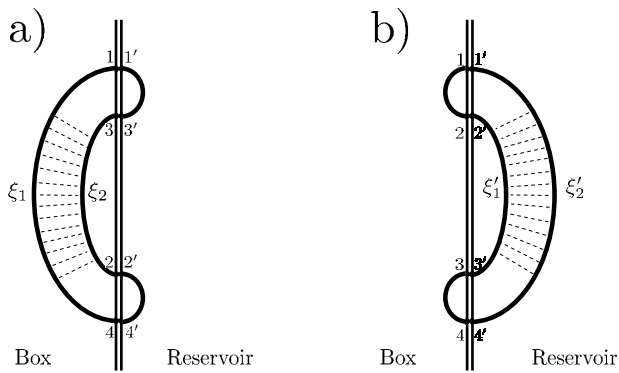}
\caption{ The diagrams corresponding to the interference of
electron trajectories in the box ($a$) and reservoir ($b$). The
contribution of the diagrams with interference in both electrodes
(not shown) is much smaller than the one of the above
diagrams~\cite{Hekking1994}.}\label{interference}
\end{figure}

It is well-known that Andreev conductance is sensitive to
disorder, see, for example, Refs.~[\onlinecite{Hekking1994,
Pothier}]. Similarly, the rate $\Gamma_{A}$ is affected by
electron backscattering to the tunnel junction,  see
Fig.~\ref{interference}.
If a quasiparticle bounces off the walls of the box or impurities
many times, it is reasonable to expect the chaotization of its
motion. Thus, one is prompted to consider ensemble-averaged
quantities rather than their particular realization. Using
Eqs.~(\ref{rate}) and~(\ref{A1}), we obtain
\begin{widetext}
\begin{eqnarray}\label{A2}
\langle\Gamma_{A}\rangle&=&\frac{\pi}{\hbar}\langle\sum_{p,p'}\int{\!\prod_{i=1..4}d\mathbf{x_i}d\mathbf{x_i'}}T(\mathbf{x_1},\mathbf{x_1'})T(\mathbf{x_2},\mathbf{x_2'})T(\mathbf{x_3},\mathbf{x_3'})T(\mathbf{x_4},\mathbf{x_4'})
\left(u_{p'}v_{p}\phi_{p'}(\mathbf{x_1'})\phi_{p}(\mathbf{x_2'})\!-\!u_{p}v_{p'}\phi_{p}(\mathbf{x_1'})\phi_{p'}(\mathbf{x_2'})\right)\nonumber\\
&\times&\left(u_{p'}v_{p}\phi_{p'}(\mathbf{x_3'})\phi_{p}(\mathbf{x_4'})\!-\!u_{p}v_{p'}\phi_{p}(\mathbf{x_3'})\phi_{p'}(\mathbf{x_4'})\right)\sum_{k}\frac{u_{k}v_{k}\phi_k(\mathbf{x_1})\phi_{k}(\mathbf{x_2})}{E_p\!+\!\delta
E_{+}\!-\!E_k}\sum_{k'}\frac{u_{k'}v_{k'}\phi_{k'}(\mathbf{x_3})\phi_{k'}(\mathbf{x_4})}{E_p\!+\!\delta
E_{+}\!-\!E_{k'}}\nonumber\\
&\times&\delta(E_{p'}-E_{p}-E_{_J})f_{F}(E_p)(1-f_F(E_{p'}))\rangle.
\end{eqnarray}
\end{widetext}
Here the brackets $\langle ... \rangle$ denote averaging
independently over different realizations of the random potential
in the box and reservoir. In order to average over the disorder in
the CPB, one has to calculate the following correlation function:
\begin{eqnarray}\label{I}
\!\!\!\!I\!&\!\!\equiv\!\!&\!\left\langle\sum_{k,k'}\frac{u_{k}v_{k}\phi_k(\mathbf{x_1})\phi_{k}(\mathbf{x_2})}{E_p\!+\!\delta
E_{+}\!-\!E_k}\frac{u_{k'}v_{k'}\phi_{k'}(\mathbf{x_3})\phi_{k'}(\mathbf{x_4})}{E_p\!+\!\delta
E_{+}\!-\!E_{k'}}\right\rangle\nonumber\\\nonumber\\
\!\!&\!\!=\!\!&\!\!\int\!\frac{\Delta_b^2{d\xi_1d\xi_2}}{4E(\xi_1)E(\xi_2)}\!\frac{\left\langle
K_{\xi_1}(\mathbf{x_1},\mathbf{x_2})
K_{\xi_2}(\mathbf{x_3},\mathbf{x_4})\right\rangle}{[E_p\!+\!\delta
E_{+}\!-\!E(\xi_1)][E_p\!+\!\delta
E_{+}\!-\!E(\xi_2)]}, \nonumber\\
\end{eqnarray}
where
$K_{\xi}(\mathbf{x_1},\mathbf{x_2})=\sum_{k}\phi_k(\mathbf{x_1})\phi_{k}(\mathbf{x_2})\delta(\epsilon_k\!-\xi)$,
and $E(\xi)=\sqrt{\xi^2+\Delta_b^2}$. The correlation function
$\langle K_{\xi_1}(\mathbf{x_1},\mathbf{x_2})
K_{\xi_2}(\mathbf{x_3},\mathbf{x_4})\rangle$ consists of
reducible and irreducible parts,
\begin{eqnarray}\label{KK2}
\langle
K_{\xi_1}(\mathbf{x_1},\mathbf{x_2})\!\!\!\!\!\!&\!\!\!\!\,\!\!\!\!&\!\!\!\!\!\!K_{\xi_2}(\mathbf{x_3},\mathbf{x_4})
\rangle\!=\!\langle K_{\xi_1}(\mathbf{x_1},\mathbf{x_2})\rangle
\langle
K_{\xi_2}(\mathbf{x_3},\mathbf{x_4})\rangle+\nonumber\\
\!&+&\!\langle K_{\xi_1}(\mathbf{x_1},\mathbf{x_2})
K_{\xi_2}(\mathbf{x_3},\mathbf{x_4}) \rangle_{\rm{ir}}.
\end{eqnarray}
The reducible part can be easily calculated by relating $\langle
K_{\xi}(\mathbf{x_1},\mathbf{x_2})\rangle$ to the
ensemble-averaged Green function: $\!\langle
K_{\xi}(\mathbf{x_1},\mathbf{x_2})\rangle\equiv\!-\frac{1}{\pi}{\rm{Im}}\langle
G^R_{\xi}(\mathbf{x_1},\mathbf{x_2})\rangle\!=\!\nu_F f_{12}$.
(Upon averaging over disorder, one can neglect the energy
dependence of the density of states here, \emph{i.e.}
$\langle\nu_F(\xi)\rangle=\nu_F$. The function $f_{12}$ is given
by $f_{12}=\langle
e^{i\mathbf{k}(\mathbf{\mathbf{x_1}}\!-\!\mathbf{x_2})}\rangle_{\rm{FS}}$
with $\langle...\rangle_{\rm{FS}}$ being the average over electron
momentum on the Fermi surface. For 3D system the function $f_{12}$
is equal to
$f_{12}=~\frac{\sin(k_F|\mathbf{x_1}\!-\!\mathbf{x_2}|)}{k_F|\mathbf{x_1}\!-\!\mathbf{x_2}|}$.)
The irreducible part $\langle K_{\xi_1}(\mathbf{x_1},\mathbf{x_2})
K_{\xi_2}(\mathbf{x_3},\mathbf{x_4}) \rangle_{\rm{ir}}$ can be
expressed in terms of the classical diffusion propagators
 - diffusons and Cooperons, see, for example, Aleiner \emph{et. al.}~[\onlinecite{Aleiner2002}]. In the absence of magnetic field, diffusons and Cooperons
coincide,
$\mathcal{P}_{\omega}(\mathbf{x}_1,\mathbf{x}_2)=\mathcal{P}^{D}_{\omega}(\mathbf{x}_1,\mathbf{x}_2)=\mathcal{P}^{C}_{\omega}(\mathbf{x}_1,\mathbf{x}_2)$,
and the irreducible part of the correlation function~(\ref{KK2})
reads
\begin{eqnarray}\label{KK3} \!\!\!\!\!\!&\,&\!\!\!\!\!\!\!\!\!\!\langle
K_{\xi_1}(\mathbf{x_1},\mathbf{x_2})
K_{\xi_2}(\mathbf{x_3},\mathbf{x_4})
\rangle_{\rm{ir}}=\\
\!\!&\!=\!&\!\!\frac{\nu_F}{\pi}\mbox{Re}\left[f_{14}f_{23}\mathcal{P}_{|\xi_2\!-\!\xi_1\!|}(\mathbf{x}_1,\mathbf{x}_3)\!+\!f_{13}f_{24}\mathcal{P}_{|\xi_2\!-\!\xi_1\!|}(\mathbf{x}_1,\mathbf{x}_4)\right].\nonumber
\end{eqnarray}
The spectral expansion of
$\mathcal{P}_{\omega}(\mathbf{x}_1,\mathbf{x}_2)$ for the
diffusive system is
\begin{eqnarray}
\mathcal{P}_{\omega}(\mathbf{x}_1,\mathbf{x}_2)=\sum_{n}\frac{f^*_{n}(\mathbf{x}_1)f_{n}(\mathbf{x}_2)}{-i\omega+\gamma_{n}}.
\end{eqnarray}
Here $\gamma_n$ and $f_{n}(\mathbf{x})$ are the corresponding
eigenvalues and eigenfunctions of the diffusion equation,
$-D\vec{\nabla}^2f_{n}(\mathbf{x})=\gamma_nf_{n}(\mathbf{x})$,
satisfying von Neumann boundary conditions in the box.

\begin{figure}[!htb]
\centering
\includegraphics[width=2.5in]{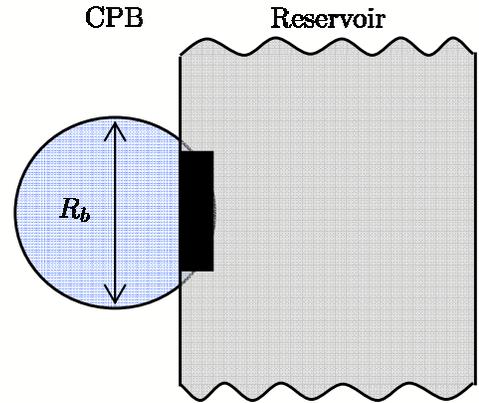}
\caption{ The layout of the Cooper-pair box qubit considered in
the text.}\label{layout}
\end{figure}

\begin{figure}
\centering
\includegraphics[width=3.2in]{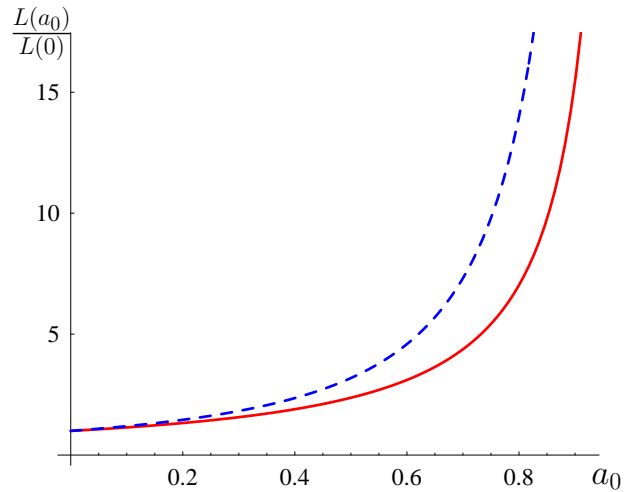}
\caption{The dependence of the functions $L_1(a_0)$ and $L_2(a_0)$
(normalized by $L_1(0)$ and $L_2(0)$, respectively) on the
dimensionless parameter $a_0=(\Delta_r\!+\!\delta
E_{+})/\Delta_b$. The solid and dashed lines correspond to $L_1$
and $L_2$, respectively, and reflect the increase of the rates
$\Gamma_1$ and $\Gamma_2$ with $a_0$. The expressions for
$L_1(a_0)$ and $L_2(a_0)$ given by Eq.~(\ref{L12}) are valid for
$a_0\ll 1-T/\Delta_b$.}\label{figL12}
\end{figure}

Equation~(\ref{KK3}) can be simplified in the case of large
Thouless energy, \emph{i.e.} $E_{_T}\gg \Delta_b, \Delta_r, E_c,
E_{_J}$. (Here $E_{_T}=\hbar/\tau_{D}$ with $\tau_{D}\sim S_b/D$
being the time to diffuse through the box, and $S_b$ being the
area of the island, see Fig.~\ref{layout}.) This condition is
fulfilled for a small aluminum island~\cite{footnote} with $S_b\ll
1\mu m ^2$ and mean free path $l\gtrsim 25nm$~\cite{Santhanam},
when the time spent by the virtual quasiparticle in the box,
$t\sim\hbar/(\Delta_b-\Delta_r-\delta E_{+})$, is much longer than
the classical diffusion time $\tau_D$~\cite{Averin}. In this case
the irreducible part in Eq.~(\ref{KK2}) is given by the universal
limit,
\begin{eqnarray}\label{KK1}
\langle K_{\xi_1}(\mathbf{x_1},\mathbf{x_2})K_{\xi_2}(\mathbf{x_3},\mathbf{x_4})\!\!\!\!\!\!&\!\!\!\!\,\!\!\!\!&\!\!\!\!\!\!\rangle_{\rm{ir}}\\
\!&=&\!\frac{\nu_F}{V_b}
\delta(\xi_1-\xi_2)\!\left(\!f_{14}f_{23}\!+\!f_{13}f_{24}\right)\!.\nonumber
\end{eqnarray}
Here $V_b$ is the volume of the box. Upon substituting
Eqs.~(\ref{KK2}) and (\ref{KK1}) into Eq.~(\ref{I}) and evaluating
the integrals over energies $\xi_1$ and $\xi_2$, we obtain
\begin{eqnarray}\label{I2}
I&=&4\nu_F^2f_{12}f_{34}L_1\!\left[\frac{E_p\!+\!\delta
E_{+}}{\Delta_b}\right]\nonumber\\
\!&+&\!\nu_F^2\frac{\delta_b}{2\Delta_b}\left(f_{14}f_{23}\!+\!f_{13}f_{24}\right)L_2\!\left[\frac{E_p\!+\!\delta
E_{+}}{\Delta_b}\right],
\end{eqnarray}
where $\delta_b=1/\nu_F V_b$ is mean level spacing in the box. The
functions $L_1(y)$ and $L_2(y)$ are defined as
\begin{eqnarray}\label{L12}
L_1(y)&=&\frac{1}{1\!-\!y^2}\arctan^2\left(\sqrt{\frac{1\!+\!y}{1\!-\!y}}\right),\nonumber\\
L_2(y)&=&\int_{1}^{\infty}{d
x}\frac{1}{\sqrt{x^2\!-\!1}}\frac{1}{x(x\!-\!y)^2}.
\end{eqnarray}
The expressions above are valid for $y<1$. The function $L_2(y)$
has the following asymptotes
\begin{eqnarray}\label{asymptotes}
L_2(y)\!\approx\!\left\{
\begin{array}{rcl}
&\!&\frac{\pi}{4}+\frac{4}{3}y,
\,\,\,\,\,\,\,\,\,\,\,\,\,\,\,\,\,\,\,\,\,\,y\ll 1
, \\\\
&\!&\frac{\pi}{2\sqrt{2}(1-y)^{3/2}},\,\!\!\!\,\,\,\,\, 1\!-\!y\ll 1.\\
\end{array}
\right.
\end{eqnarray}
After substituting Eq.~(\ref{I2}) into Eq.~(\ref{A2}) and
averaging over disorder in the reservoir, we obtain the following
expression for $\langle\Gamma_{A}\rangle$:
\begin{widetext}
\begin{eqnarray}\label{rate2}
\!\!\!\langle\Gamma_{A}\rangle\!\!&\!\!=\!\!&\!\!\frac{\pi\nu_F^2}{2\hbar}\int{d\xi'_1d\xi'_2}\delta(E(\xi'_2)\!-\!E(\xi'_1)\!-\!E_{_J})f_{F}[E(\xi'_1)]\!\left(1\!-\!f_{F}[E(\xi'_2)]\right)\!\int\!{\!\prod_{i=1..4}d\mathbf{x}_id\mathbf{x}_i'}
T(\mathbf{x_1},\mathbf{x_1'})T(\mathbf{x_2},\mathbf{x_2'})T(\mathbf{x_3},\mathbf{x_3'})T(\mathbf{x_4},\mathbf{x_4'})\nonumber\\
\!\!&\!\!\times\!\!&\!\!\left(\!4f_{12}f_{34}L_1\!\left[\!\frac{E(\xi'_1)\!+\!\delta
E_{+}}{\Delta_b}\!\right]\!+\!\frac{\delta_b}{2\Delta_b}\left(f_{14}f_{23}\!+\!f_{13}f_{24}\right)L_2\!\left[\!\frac{E(\xi'_1)\!+\!\delta
E_{+}}{\Delta_b}\!\right]\!\right)\!\left(\!1\!-\!\frac{\Delta_r^2}{E(\xi'_1)E(\xi'_2)}\!\right)\!\langle K_{\xi'_1}(\mathbf{x'}_1,\mathbf{x'}_3) K_{\xi'_2}(\mathbf{x'}_2,\mathbf{x'}_4) \rangle.\nonumber\\
\end{eqnarray}
\end{widetext}
Here $E(\xi')=\sqrt{\xi'^2+\Delta_r^2}$. The correlation function
in the reservoir $\langle K_{\xi'_1}(\mathbf{x'}_1,\mathbf{x'}_3)
K_{\xi'_2}(\mathbf{x'}_2,\mathbf{x'}_4) \rangle$ follows from
Eqs.~(\ref{KK2}) and~(\ref{KK3}). Using Eq.~(\ref{tunn}) and
evaluating the spatial integrals over the area of the junction as
well as the integrals over energies $\xi'_1$, and $\xi'_2$, we
finally obtain the answer for $\langle\Gamma_{A}\rangle$:
\begin{eqnarray}\label{rate4}
\langle\Gamma_{A}\rangle &=& \Gamma_1+\Gamma_2
\end{eqnarray}
with $\Gamma_1$ and $\Gamma_2$ being defined as
\begin{eqnarray}\label{Gamma1}
\Gamma_1 \!\approx\!
\frac{2\pi}{\hbar}\frac{3C_1}{(4\pi^2)^2}\frac{g_{_T}^2}{N_{\rm{ch}}}\sqrt{\frac{E_{_J}}{2\Delta_r\!+\!E_{_J}}}\frac{n_{\rm{qp}}}{\nu_F}L_1\!\left[\frac{\Delta_r\!+\!\delta
E_{+}}{\Delta_b}\right]\!,
\end{eqnarray}
and
\begin{eqnarray}\label{Gamma2}
\Gamma_2 \!\approx\!
\frac{2\pi}{\hbar}\frac{g_{_T}^2}{8(4\pi^2)^2}\frac{\delta_b}{\Delta_b}\sqrt{\frac{E_{_J}}{2\Delta_r\!+\!E_{_J}}}\frac{n_{\rm{qp}}}{\nu_F}
 L_2\!\left[\frac{\Delta_r\!+\!\delta E_{+}}{\Delta_b}\right].
\end{eqnarray}
Here $C_1$ is a numerical constant of the order of one:
\begin{eqnarray}
C_1=\frac{1}{\pi^3
k_F^2S_{_J}}\int_{k_F^2S_{_J}}{d\mathbf{y}_1d\mathbf{y}_2d\mathbf{y}_3d\mathbf{y}_4}P_{12}P_{13}P_{24}P_{34}\nonumber
\end{eqnarray}
with $\mathbf{y}$ being a dimensionless coordinate in the plane of
a tunnel junction, and
$P_{12}=\frac{\sin(|\mathbf{y}_1-\mathbf{y}_2|)-|\mathbf{y}_1-\mathbf{y}_2|\cos(|\mathbf{y}_1-\mathbf{y}_2|)}{|\mathbf{y}_1-\mathbf{y}_2|^3}$.
The functions $L_1$ and $L_2$ are defined in Eq.~(\ref{L12}), and
their dependence on the ratio $(\Delta_r\!+\!\delta
E_{+})/\Delta_b$ is shown in Fig.~\ref{figL12}. The rate
$\Gamma_1$ describes the contribution from the reducible terms,
see Eq.~(\ref{KK2}), and is similar to the ballistic case when
electron scattering from the impurities or boundaries is
negligible. The other term, $\Gamma_2$, reflects the enhancement
of $\langle\Gamma_{A}\rangle$ in the diffusive limit due to the
quantum interference of quasiparticle return
trajectories~\cite{magnetic}, and originates from the irreducible
contributions, see Fig.~\ref{interference}. In the case of
$N_{\rm{ch}}\delta_b/\Delta_b \gg 1$, the contribution of this
interference term becomes dominant, $\Gamma_2\gg\Gamma_1$. The
contribution of the interference in the reservoir to the rate
$\Gamma_2$, see Fig.~\ref{interference}b, is geometry dependent.
For a typical charge qubit with the small junction connected to a
large electrode, backscattering of electrons to the junction from
the reservoir side gives much smaller contribution to $\Gamma_2$
than the similar one for the box side of the junction. In
particular, for the layout of the qubit shown in
Fig.~\ref{layout}, the contribution of the interference in the
reservoir to $\Gamma_2$ is smaller than the one in the box by a
factor $\frac{d_b}{d_r}\frac{\Delta_b}{E_{_T}}\ln\left[\frac{\hbar
D}{\Delta_rS_{J}}\right]\frac{L_1(a_0)}{L_2(a_0)}\ll 1$. [Here
$a_0=(\Delta_r\!+\!\delta E_{+})/\Delta_b$, and $d_{b(r)}$ is the
thickness of the superconducting film in the box(reservoir).]
Therefore, we neglected the terms corresponding to the
interference in the reservoir in Eq.~(\ref{Gamma2}).

\section{Conclusion}
We have studied the fundamental limitations on the energy
relaxation time in a charge qubit with a large-gap Cooper-pair
box, $\Delta_b> \Delta_r$. For sufficiently large $\Delta_b$, real
quasiparticle transitions can be exponentially suppressed, and the
dominant contribution to the charge qubit energy relaxation time
$T_1$ comes from the higher-order (Andreev) processes, see
Eq.~(\ref{rate4}). For realistic geometry of the charge qubits and
the density of nonequilibrium quasiparticles in the reservoir
$n_{\rm qp}\sim
10^{19}-10^{18}\rm{m^{-3}}$~[\onlinecite{Lutchyn2006}], we
estimate the Andreev relaxation rate to be
$\langle\Gamma_{A}\rangle\sim 10^{-1}-10^{-2}$Hz. Thus, in the
absence of other relaxation channels, the mismatch of gap energies
leads to extremely long $T_1$-times. (For comparison, the
quasiparticle-induced
 $T_1$ found in Ref.~[\onlinecite{Lutchyn2006}] for the charge qubit with equal
gap energies  was $T_1^{-1}\sim 10^5-10^3$Hz.)

The charge qubit with a large gap in the box also permits to
reduce quasiparticle-induced decoherence. Since real quasiparticle
transitions into the island are suppressed, see
Eq.~(\ref{T1_exp}), the dephasing time of the qubit is limited by
the energy relaxation processes, \emph{i.e.} $T_2\approx
2/\langle\Gamma_{A}\rangle$.

\begin{acknowledgments}
This work was supported by NSF grants DMR 02-37296, and DMR
04-39026.
\end{acknowledgments}


\begin{thebibliography}{99}
\bibitem{footnote} For an island shown in Fig.~\ref{layout},
$E_{_T}=14.7\pi\hbar D/S_b$. The superconducting gap $\Delta_b$
can be expressed in terms of the diffusion constant $D$ and the
coherence length in a system with disorder $\xi_{dirty}$:
$\Delta_b=\frac{1.74}{2\pi}\hbar D/\xi^2_{dirty}$. Thus, the ratio
$\Delta_b/E_{_T}\approx 6\cdot 10^{-3} S_b/\xi^2_{dirty}\ll 1$
sets constraints on the size of the box.
\end{thebibliography}
\end{document}